\documentclass[preprint,showpacs,preprintnumbers,amsmath,amssymb]{revtex4}

\usepackage{amssymb}
\usepackage{amsmath}
\usepackage{graphicx}
\usepackage{epsfig}
\usepackage{subfigure}
\usepackage{amsfonts}
\usepackage{CJK}

\begin{document}

\title{Efficient scheme for generation of photonic NOON states in circuit QED}

\author{Shao-Jie Xiong$^{1}$, Zhe Sun$^{1}$, Jin-Ming Liu$^{2}$, Tong Liu$^{1}$, and Chui-Ping Yang$^{1}$}

\address{$^1$Department of Physics, Hangzhou Normal University,
Hangzhou, Zhejiang 310036, China}

\address{$^2$State Key Laboratory of Precision Spectroscopy, East China Normal University,
Shanghai 200062, China}
\date{\today}

\begin{abstract}
We propose an efficient scheme for generating photonic NOON states of two resonators
coupled to a four-level superconducting flux device. This proposal operates
essentially by employing a technique of a coupler device resonantly
interacting with two resonators simultaneously. As a consequence, the NOON-state
preparation requires only $N+1$ operational steps and thus is much faster when
compared with a recent proposal [Q. P. Su et al., Scientific Reports 4, 3898 (2014)]
requiring $2N$ steps of operation. Moreover, due to the use of only two resonators and a device,
the experimental setup is much simplified when compared with previous proposals
requiring three resonators and two superconducting qubits/qutrits.
\end{abstract}

 \maketitle
\date{\today}

In recent years there is considerable interest in the entangled NOON states $%
\left| N0\right\rangle +\left| 0N\right\rangle $, which have significant
applications in quantum optical lithography [1], quantum metrology [2],
precision measurement of transmons [3], and quantum information processing
[4]. Based on circuit QED [5,6], several proposals have been presented to
generate the photonic NOON states of two resonators [7-11].

The scheme in [7] requires that the pulse Rabi frequency is much
smaller than the photon-number-dependent Stark shifts induced by dispersive
interaction. Thus, the operation time needed to complete a rotation in each
step is quite long. Another method was proposed [8] and implemented in
experiment for $N\leq 3$ with a fidelity $0.33$ for $N=3$ [9]. This method
shortens the operation time due to using resonant interaction but needs
a complex setup (i.e., three resonators and two superconducting qutrits),
which increases the experimental difficulty. Moreover, two classical pulses
(e.g., a double pulse) are separately applied to the two qutrits
during each step, conditional on the NOON state being prepared with $N$
steps. The scheme in [10] employs a complicated pulse. Similar to
[8,9], this scheme requires two auxiliary superconducting qubits initially
prepared in a Bell state. As argued there, to obtain a pure photonic NOON
state, additional techniques are required to decouple the qubits from the
resonators.

Recently, Q.P.Su \textit{et al.} have proposed an alternative scheme for
generating the NOON states of two resonators or cavities [11]. Compared with
the previous proposals [8-10], the experimental setup is greatly simplified
because of employing one superconducting qutrit and two resonators only. Due
to using the resonant interaction, the operation can be performed much
faster when compared with the method in [7]. However, as discussed in [11], $%
2N$ steps of operation are needed.

We here employ a four-level superconducting flux device to couple two
resonators (hereafter the term cavity and resonator is used
interchangeably). Different from the previous proposals [7-11], the device
is simultaneously resonant with two cavities and thus two photons can be
simultaneously created each in one cavity for each of the first $N-1$
operational steps.

This scheme only requires $N+1$ operational steps and thus the operation is
much speeded up when compared with the recent proposal [11] requiring $2N$
steps. Numerical simulation shows that a high fidelity generation of the
NOON state with $N\leq 10$ is feasible within present-day circuit QED.
Further, this scheme has additional advantages: (i) Because of using only
two resonators and a device, the setup is much simplified when compared with
Refs.~[8-10]; (ii) Due to using resonant interaction, the operation can be
performed much faster when compared with [7]. Hence, the present scheme
avoids most of the problems existing in the previous proposals.

\begin{figure}[tbp]
\begin{center}
\includegraphics[bb=22 20 595 291, width=5 cm, height=3 cm,clip]{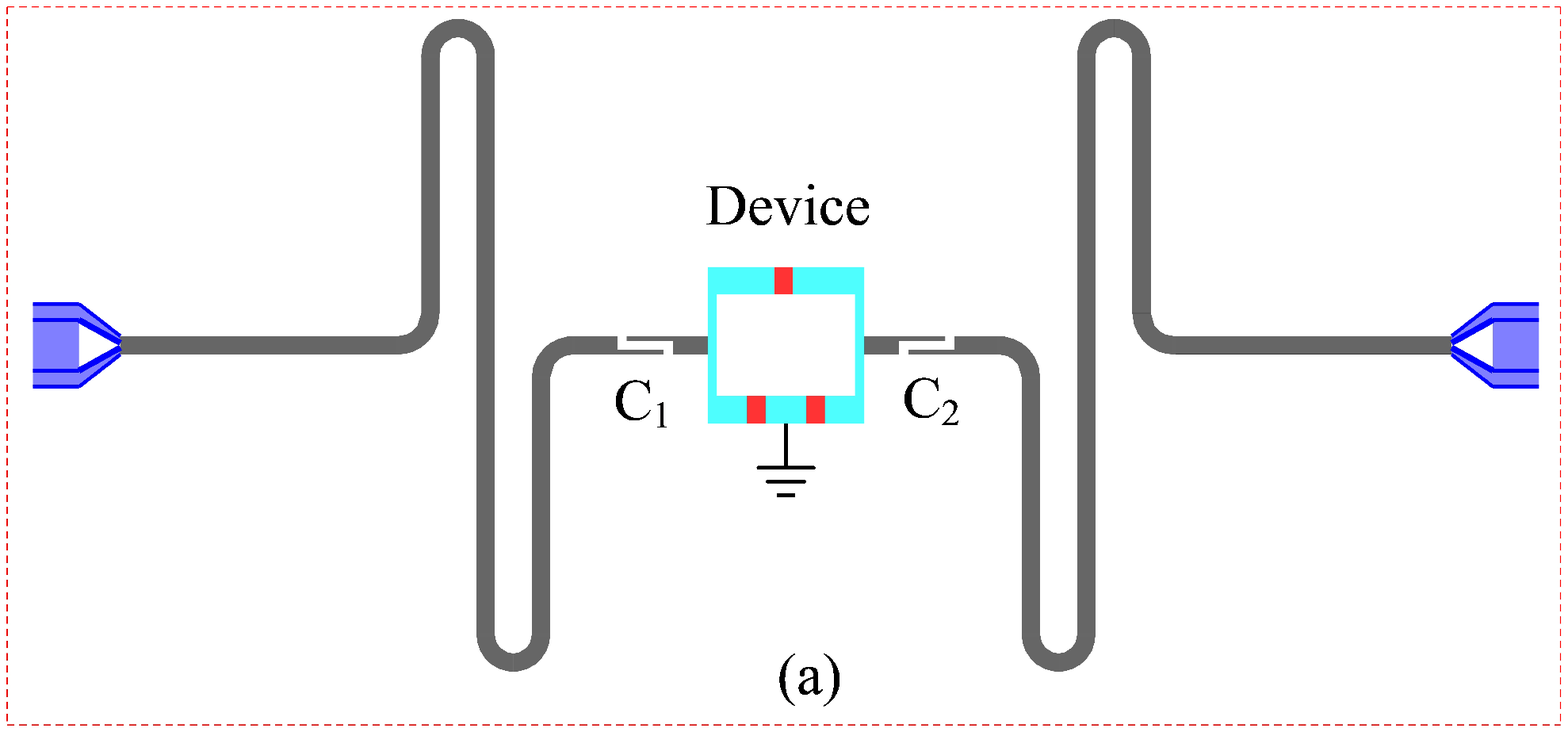} %
\includegraphics[bb=27 16 590 611, width=2.5 cm, height=3 cm,clip]{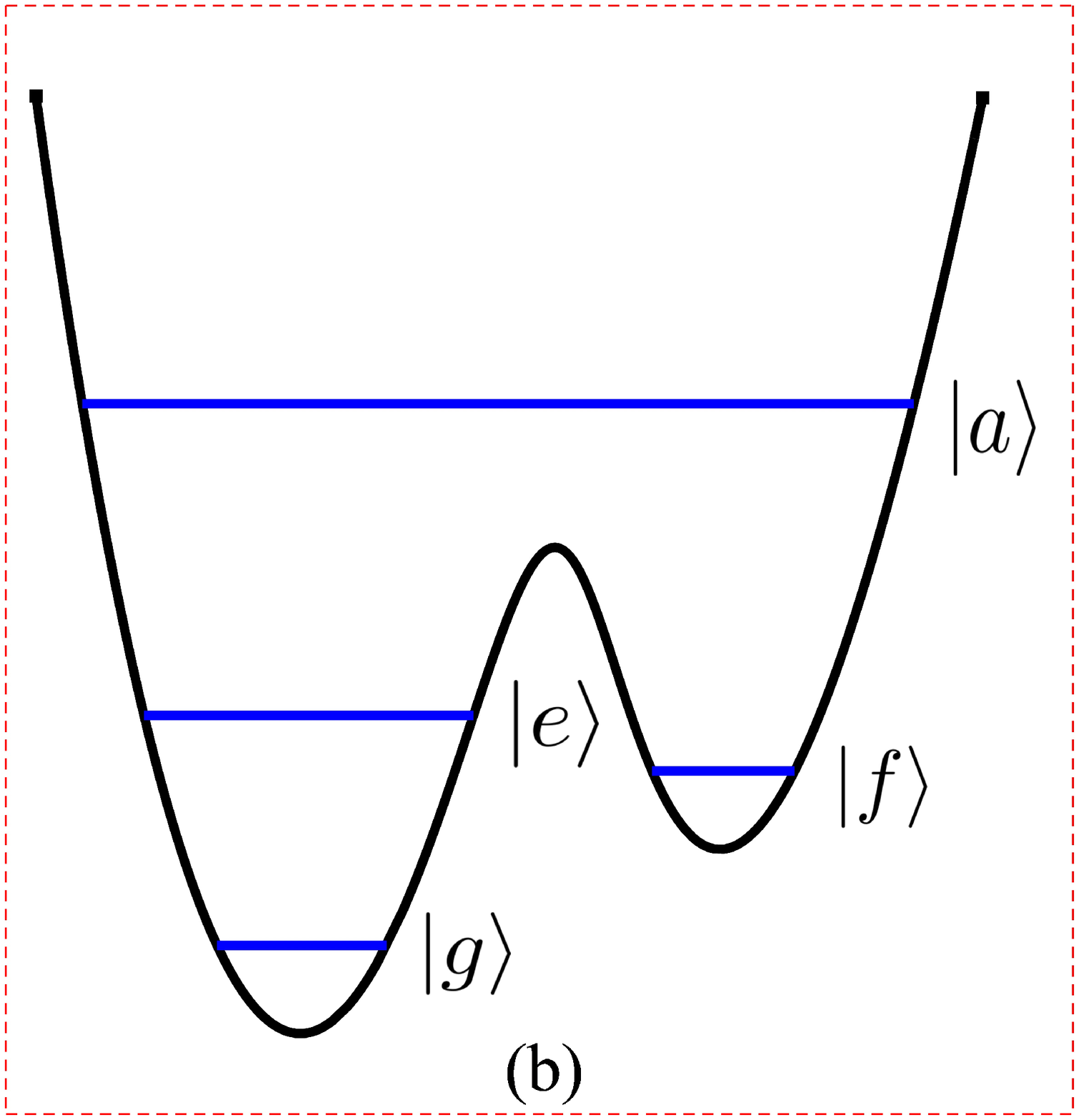}
\vspace*{-0.08in}
\end{center}
\caption{(color online). (a) Setup for two cavities and a superconducting
flux device. Each cavity is a one-dimensional transmission
line resonator. The device is connected to the two cavities via capacitors $%
c_i(i=1,2)$. (b) Four Levels of the device with a double potential well.}
\label{fig:1}
\end{figure}

Consider two cavities coupled to a flux device with four
levels $|g\rangle ,~|e\rangle ,~|f\rangle ,~$and $|a\rangle $ (Fig.~1).
Initially, the device is in the state $\frac{1}{\sqrt{2}}(\left\vert
e\right\rangle +\left\vert a\right\rangle )$ and each cavity in a vacuum
state $|0\rangle $. The device is initially decoupled from the two cavities,
which can be achieved by a prior adjustment of the device level spacings.
Note that for a flux device, the level spacings can be rapidly adjusted via
varying external control parameters [12,13]).

Define $\omega _{eg},$ $\omega _{af},$ $\omega _{ae}$ as the $|g\rangle
\leftrightarrow |e\rangle ,$ $|f\rangle \leftrightarrow |a\rangle $ and $%
|e\rangle \leftrightarrow |a\rangle $ transition frequencies of the device,
respectively. The frequency, initial phase, and duration of the pulse are
denoted as \{$\omega ,$ $\varphi ,$ $t$\}.
\begin{figure}[tbp]
\begin{center}
\includegraphics[bb=19 16 595 263, width=8 cm, height=3 cm,clip]{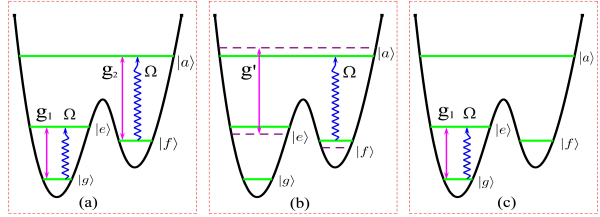}
\vspace*{-0.08in}
\end{center}
\caption{(color online). Illustration of resonant interaction between the
device and the cavity/pulse during the NOON state preparation. Figures (a),
(b) and (c) are for first $N-1$ steps, step $N$, and step $N+1$,
respectively. In (b), dashed lines represent the adjusted energy levels.}
\label{fig:2}
\end{figure}

To begin with, the level spacings of the device needs to be adjusted to have
cavity $1$ ($2$) resonant with $\left\vert g\right\rangle
\leftrightarrow \left\vert e\right\rangle $ ($\left\vert f\right\rangle
\leftrightarrow \left\vert a\right\rangle $) transition [Fig.~2(a)]. The
procedure for the NOON-state generation is described below:

Step $1$: Let cavity 1 (2) resonant with the $|g\rangle \leftrightarrow
|e\rangle $ ($|f\rangle \leftrightarrow |a\rangle $) transition [Fig.~2
(a)]. In the interaction picture (the same picture is used without
mentioning hereafter), the interaction Hamiltonian is $H_{1}=g_{1}(a_{1}%
\sigma _{eg}^{+}+h.c.)+g_{2}(a_{2}\sigma _{af}^{+}+h.c.),$ where $\sigma
_{eg}^{+}=|e\rangle \langle g|$, $\sigma _{af}^{+}=|a\rangle \langle f|,$ $%
a_{1}$ ($a_{2}$) is the photon annihilation operator for cavity $1$ ($2$),
and $g_{1}$ ($g_{2}$) is the coupling strength between cavity $1$ ($2$)\ and
the $|g\rangle \leftrightarrow |e\rangle $ ($|f\rangle \leftrightarrow
|a\rangle $) transition. Note that $g_{1}$ ($g_{2}$) depends on the coupling
capacitance $c_{1}$ ($c_{2}$). Thus, set $g_{1}=g_{2}=g$, which can in
principle be met by a prior design of the sample with appropriate $c_{1}$
and $c_{2}.$ Under $H_{1}$, the time evolution of the states $\left\vert
e\right\rangle \left\vert n\right\rangle _{1}\left\vert 0\right\rangle _{2}$
and $\left\vert a\right\rangle \left\vert 0\right\rangle _{1}\left\vert
n\right\rangle _{2}$ is described by
\begin{eqnarray}
\left\vert e\right\rangle \left\vert n\right\rangle _{1}\left\vert
0\right\rangle _{2} &\rightarrow &C\left\vert e\right\rangle \left\vert
n\right\rangle _{1}\left\vert 0\right\rangle _{2}-iD\left\vert
g\right\rangle \left\vert n+1\right\rangle _{1}\left\vert 0\right\rangle
_{2}, \\
\left\vert a\right\rangle \left\vert 0\right\rangle _{1}\left\vert
n\right\rangle _{2} &\rightarrow &C\left\vert a\right\rangle \left\vert
0\right\rangle _{1}\left\vert n\right\rangle _{2}-iD\left\vert
f\right\rangle \left\vert 0\right\rangle _{1}\left\vert n+1\right\rangle
_{2},
\end{eqnarray}%
where $C=\cos (\sqrt{n+1}gt),$ $D=\sin (\sqrt{n+1}gt),$ subscripts $1$ ($2$)
represents cavity 1 (2), $\left\vert n\right\rangle $ and $\left\vert
n+1\right\rangle $ are the cavity photon-number states. For simplicity,
define $|i\rangle _{1}|j\rangle _{2}=$ $|i\rangle |j\rangle $ with $i,j\in
\{0,1,...,N\}$. Eqs.~(1) and (2) show that after an interaction time $t_{1}=%
\frac{\pi }{2g}$(i.e., half a Rabi oscillation)$,$ the state $|e\rangle
|0\rangle |0\rangle $ changes to $-i|g\rangle |1\rangle |0\rangle $ while
the state $|a\rangle |0\rangle |0\rangle $ changes to $-i|f\rangle |0\rangle
|1\rangle .$ Thus, the initial state $\frac{\sqrt{2}}{2}(|e\rangle
+|a\rangle )|0\rangle |0\rangle $ of the system evolves to
\begin{equation}
\frac{-i}{\sqrt{2}}(|g\rangle |1\rangle |0\rangle +|f\rangle |0\rangle
|1\rangle ).
\end{equation}%
Now, apply a double pulse of $\{\omega _{eg},-\frac{\pi }{2},\frac{\pi }{%
2\Omega }\}$ and $\{\omega _{af},-\frac{\pi }{2},\frac{\pi }{2\Omega }\}$ to
the device. The identical Rabi frequency $\Omega $ of each pulse can be
achieved by adjusting the pulse intensities. The interaction Hamiltonian is
given by $H_{2}=(\Omega e^{i\pi /2}\sigma _{eg}^{+}+h.c.)+(\Omega e^{i\pi
/2}\sigma _{af}^{+}+h.c.),$ Under the Hamiltonian $H_{2}$, the time
evolution of the states $\left\vert g\right\rangle ,$ $\left\vert
e\right\rangle ,$ and $\left\vert f\right\rangle $ is described by

\begin{eqnarray}
\left\vert g\right\rangle  &\rightarrow &\cos \Omega t\left\vert
g\right\rangle +\sin \Omega t\left\vert e\right\rangle , \\
\left\vert e\right\rangle  &\rightarrow &-\sin \Omega t\left\vert
g\right\rangle +\cos \Omega t\left\vert e\right\rangle , \\
\left\vert f\right\rangle  &\rightarrow &\cos \Omega t\left\vert
f\right\rangle +\sin \Omega t\left\vert a\right\rangle .
\end{eqnarray}%
For $\Omega \gg g$, the interaction between cavities and the device can be
neglected during the pulse. Based on Eqs.~(4-6), the pulse leads to $%
|g\rangle \rightarrow \left\vert e\right\rangle $ and $|f\rangle \rightarrow
|a\rangle .$ As a consequence, the state~(3) becomes
\begin{equation}
\frac{-i}{\sqrt{2}}(|e\rangle |1\rangle |0\rangle +|a\rangle |0\rangle
|1\rangle ).
\end{equation}

Step $j$ ($j=2,3,...,N-1$): Repeat the operation of step 1. The time for the
device resonant with the two cavities is $t_{j}=\frac{\pi }{2\sqrt{j}g}$.
Eqs.~(1) and (2) show that after $t_{j},$ the state $|e\rangle |j-1\rangle
|0\rangle $ changes to $-i|g\rangle |j\rangle |0\rangle $ and the state $%
|a\rangle |0\rangle |j-1\rangle $ becomes $-i|f\rangle |0\rangle |j\rangle $%
, which further turn into $-i|e\rangle |j\rangle |0\rangle $ and $%
-i|a\rangle |0\rangle |j\rangle $ respectively, due to a double pulse of $%
\{\omega _{eg},-\frac{\pi }{2},\frac{\pi }{2\Omega }\}$ and $\{\omega _{af},-%
\frac{\pi }{2},\frac{\pi }{2\Omega }\}$ pumping the state $\left\vert
g\right\rangle $ back to $\left\vert e\right\rangle $ and the state $%
\left\vert f\right\rangle $ back to $\left\vert a\right\rangle $. Hence,
after step $N-1,$ the state (7) changes to
\begin{equation}
\frac{(-i)^{N-1}}{\sqrt{2}}(|g\rangle |N-1\rangle |0\rangle +|f\rangle
|0\rangle |N-1\rangle ).
\end{equation}

Step $N$: Apply a pulse of $\{\omega _{af},-\frac{\pi }{2},\frac{\pi }{%
2\Omega }\}$ to the device [Fig.~2(b)], described by $H_{3}=\Omega e^{i\pi
/2}\sigma _{af}^{+}+h.c.,$ i.e., the second term in $H_{2}$. Note that the
first term $(\Omega e^{i\pi /2}\sigma _{eg}^{+}+h.c.)$ in $H_{2}$ acting on
the state $\left\vert f\right\rangle $ or $\left\vert a\right\rangle $
equals to zero. Thus, under the Hamiltonian $H_{3}$, the time evolution of
the state $\left\vert f\right\rangle $ is given by Eq.~(6), which shows that
after the pulse, the state $\left\vert f\right\rangle $ changes to $%
\left\vert a\right\rangle $. Thus, the state $|f\rangle \left\vert
0\right\rangle |N-1\rangle $ changes to $|a\rangle \left\vert 0\right\rangle
|N-1\rangle .$ Now, tune the level spacing of the device so that cavity 1 is
decoupled from the device but cavity 2 resonant with the $|e\rangle
\leftrightarrow |a\rangle $ transition [Fig.~2(b)]. The interaction
Hamiltonian is $H_{4}=g^{\prime }(a_{2}^{\dagger }\sigma _{ae}^{-}+h.c.),$
where $\sigma _{ae}^{-}=\left\vert e\right\rangle \left\langle a\right\vert $%
, $g^{\prime }$ is the coupling constant between cavity $2$ and $|e\rangle
\leftrightarrow |a\rangle $ transition. The state time evolution is
described by
\begin{equation}
\left\vert a\right\rangle \left\vert 0\right\rangle \left\vert
n\right\rangle \rightarrow C^{\prime }\left\vert a\right\rangle \left\vert
0\right\rangle \left\vert n\right\rangle -iD^{\prime }\left\vert
e\right\rangle \left\vert 0\right\rangle \left\vert n+1\right\rangle ,
\end{equation}%
where $C^{\prime }=\cos (\sqrt{n+1}g^{\prime }t),$ $D^{\prime }=\sin (\sqrt{%
n+1}g^{\prime }t).$ According to Eq.~(9), the state $|a\rangle |0\rangle
|N-1\rangle $ becomes $-i|e\rangle |0\rangle |N\rangle $ after an
interaction time $t_{N}=\frac{\pi }{2\sqrt{N}g^{\prime }}$, but the state $%
|g\rangle |N-1\rangle |0\rangle $ remains unchanged due to $H_{4}|g\rangle
|N-1\rangle |0\rangle =0$. Thus, the state~(8) evolves into
\begin{equation}
\frac{(-i)^{N-1}}{\sqrt{2}}(|g\rangle |N-1\rangle |0\rangle -i|e\rangle
|0\rangle |N\rangle ).
\end{equation}

Step $N+1$: Tune the level spacing of the device back to Fig.~2(a) [i.e.,
Fig.~2(c)]. Apply a pulse of $\{\omega _{eg},-\frac{\pi }{2},\frac{\pi }{%
2\Omega }\}$ to the device [Fig.~2(c)], described by the Hamiltonian $%
H_{5}=\Omega e^{i\pi /2}\sigma _{eg}^{+}+h.c.$. It can be verified that
under $H_{5},$ the time evolution of the states $\left\vert g\right\rangle $
and $\left\vert e\right\rangle $ are given by Eqs.~(4) and (5), which show
that the transformations $|g\rangle \rightarrow |e\rangle $and $|e\rangle
\rightarrow -|g\rangle $ are obtained after the pulse. Thus, the state (10)
changes to
\begin{equation}
\frac{(-i)^{N-1}}{\sqrt{2}}(|e\rangle |N-1\rangle |0\rangle +i|g\rangle
|0\rangle |N\rangle ).
\end{equation}%
Let cavity 1 (2) resonant with the $|g\rangle \leftrightarrow |e\rangle $ ($%
|f\rangle \leftrightarrow |a\rangle $) transition for a time $t_{N+1}=\frac{%
3\pi }{2\sqrt{N}g}$. As a result, the state $|e\rangle |N-1\rangle |0\rangle
$ changes to $i|g\rangle |N\rangle |0\rangle $ according to Eq.~(1), while
the state $|g\rangle |0\rangle |N\rangle $ remains unchanged due to $%
H_{1}|g\rangle |0\rangle |N\rangle =0.$ Thus, one gets
\begin{equation}
\frac{(-i)^{N+2}}{\sqrt{2}}|g\rangle (|N\rangle |0\rangle +|0\rangle
|N\rangle ).
\end{equation}%
To maintain the state~(12), the level spacings of the device needs to be
adjusted so that the device is decoupled from the two cavities after the
entire operation. Eq.~(12) shows that the two cavities are prepared in a
NOON state and disentangled from the device.

The above description shows that no adjustment of the cavity frequencies is
needed during the entire operation. Similar to [11], the NOON-state
generation utilizes classical pulses with only two different frequencies,
readily achieved in experiment. Moreover, no measurement on the states of
the device or the two cavities is required.

The condition $g_{1}=g_{2}$ above is unnecessary. For the case of $g_{1}\neq g_{2},$ the last two steps of
operation remain the same but the first ($N-1$)-step operations are not
synchronous for subspaces I (cavity 1, $\left\vert g\right\rangle $, $%
\left\vert e\right\rangle $) and subspace II (cavity 2, $\left\vert
f\right\rangle $, $\left\vert a\right\rangle $). For instance, if $%
g_{1}>g_{2},$ the half Rabi oscillations for subspace I are completed
earlier than those of subspace II because $t_{j}=\frac{\pi }{2\sqrt{j}g_{1}}<%
\frac{\pi }{2\sqrt{j}g_{2}}$, thus the microwave pulses of the two subspaces
should be independent and asynchronous. As a result, the first ($N-1$)-step
operations on subspace I will be completed prior to those on subspace II.
Hence, after the first ($N-1$)-step operations on subspace I, one will need
to adjust the level spacings of the device to have cavity 1 decoupled from
the $\left\vert e\right\rangle \leftrightarrow \left\vert g\right\rangle $
transition, such that the time evolution of subspace I is avoided before the
first ($N-1$)-step operations on subspace II is completed. The same
reasoning applies to the case of $g_{2}>g_{1}.$

In what follows, we will give a discussion of the fidelity of the prepared
NOON state for $N\leq 10.$ As an example, we will consider $g_{1}=0.95g$ and
$g_{2}=g$. The numerical simulation is performed by following the NOON-state
procedure described previously for the homogeneous coupling constants, with
the typical operation time $t_{j}$ (depending on $g$) given there for each
of the first ($N-1$) steps of operation.

In a realistic situation there is an inter-cavity crosstalk between the two
cavities, which is described by $\varepsilon =g_{12}\left( e^{i\Delta
t}a_{1}a_{2}^{+}+h.c.\right) $, where $g_{12}$ is the coupling strength of
the two cavities and $\Delta =\omega _{a_{2}}-\omega _{a_{1}}$ is the
detuning between the two-cavity frequencies $\omega _{a_{1}}$ and $\omega
_{a_{2}}$. In addition, there is the device-cavity interaction and the
inter-cavity crosstalk during the pulses. By taking these factors into
account, it is straightforward to modify the Hamiltonians $H_{1},$ $H_{2},$ $%
H_{3},$ $H_{4},$ and $H_{5}$ (not shown to simplify the presentation).

After considering dissipation and dephasing, the system dynamics is
determined by the master equation
\begin{eqnarray}
\frac{d\rho }{dt} &=&-i\left[ H_{k},\rho \right] +\sum_{j=1}^{2}\kappa
_{a_{j}}\mathcal{L}\left[ a_{j}\right]   \notag \\
&&\ \ +\sum_{j=e,f,g}\gamma _{aj}\mathcal{L}\left[ \sigma _{aj}^{-}\right]
+\sum_{j=f,g}\gamma _{ej}\mathcal{L}\left[ \sigma _{ej}^{-}\right] +\gamma
_{fg}\mathcal{L}\left[ \sigma _{fg}^{-}\right]   \notag \\
&&\ \ +\sum_{j=a,e,f}\gamma _{\varphi ,j}\left( \sigma _{jj}\rho \sigma
_{jj}-\sigma _{jj}\rho /2-\rho \sigma _{jj}/2\right) ,
\end{eqnarray}%
where $H_{k}$ (with $k=1,2,3,4,5$) are the modified $H_{1}$ to $H_{5}$, $%
\mathcal{L}\left[ \Lambda \right] =\Lambda \rho \Lambda ^{+}-\Lambda
^{+}\Lambda \rho /2-\rho \Lambda ^{+}\Lambda /2$ (with $\Lambda
=a_{j},\sigma _{aj}^{-},\sigma _{ej}^{-},\sigma _{fg}^{-})$,\ $\sigma
_{aj}^{-}=\left\vert j\right\rangle \left\langle a\right\vert ,$ $\sigma
_{ej}^{-}=\left\vert j\right\rangle \left\langle e\right\vert ,$ $\sigma
_{fg}^{-}=\left\vert g\right\rangle \left\langle f\right\vert ,$ and $\sigma
_{jj}=\left\vert j\right\rangle \left\langle j\right\vert ;$ $\kappa _{a_{j}}
$ is the decay rate of cavity $j$;\ $\gamma _{aj}$\ is the energy relaxation
rate for the level $\left\vert a\right\rangle $\ associated with the decay
path $\left\vert a\right\rangle \rightarrow \left\vert j\right\rangle $($%
j=e,f,g$); $\gamma _{ej}$\ is for the level $\left\vert e\right\rangle $
related to the decay path $\left\vert e\right\rangle \rightarrow \left\vert
j\right\rangle $ ($j=f,g$); $\gamma _{fg}$ is for the level $\left\vert
f\right\rangle $; and $\gamma _{\varphi ,j}$ is the dephasing rate of the
level $\left\vert j\right\rangle $ ($j=a,e,f$)\textbf{.}

The fidelity of the whole operation is given by $\mathcal{F}=\sqrt{%
\left\langle \psi _{id}\right| \rho \left| \psi _{id}\right\rangle },$ where
$\left| \psi _{id}\right\rangle $ is the ideal output state given in
Eq.~(12), while $\rho $ is the final density operator of the system (i.e.,
with unwanted couplings\textbf{,} dissipation, and dephasing considered)
after the entire operation.

For a flux device, the typical transition frequency between two neighbor
levels is between 1 and 20 GHz. As an example, consider a device with
frequencies $\nu _{fg}\sim 2.5$ GHz, $\nu _{ef}\sim 1$GHz, $\nu _{eg}\sim 3.5
$ GHz, $\nu _{ae}\sim 4.5$ GHz, and $\nu _{af}\sim 5.5$ GHz. Here, $\nu
_{ij}=\omega _{ij}/\left( 2\pi \right) $. Thus, choose cavity 1 with $\omega
_{a_{1}}\sim 2\pi \times 3.5$ GHz while cavity 2 with $\omega _{a_{2}}\sim
2\pi \times 5.5$ GHz. Parameters used in the numerical simulation are: (i) $%
\gamma _{\varphi ,f}^{-1}=5$ $\mu $s$,$$\gamma _{\varphi ,e}^{-1}=1.5$ $\mu $%
s, $\gamma _{\varphi ,a}^{-1}=0.5$ $\mu $s; (ii) $\gamma _{fg}^{-1}=10$ $\mu
$s, $\gamma _{eg}^{-1}=\gamma _{ef}^{-1}=3$ $\mu $s, $\gamma
_{ae}^{-1}=\gamma _{af}^{-1}=\gamma _{ag}^{-1}=1.5$ $\mu $s [14], and (iii) $%
\kappa _{a_{1}}^{-1}=\kappa _{a_{2}}^{-1}=20$ $\mu $s.

\begin{figure}[tbp]
\begin{center}
\includegraphics[bb=0 0 961 590, width=7.0 cm, clip]{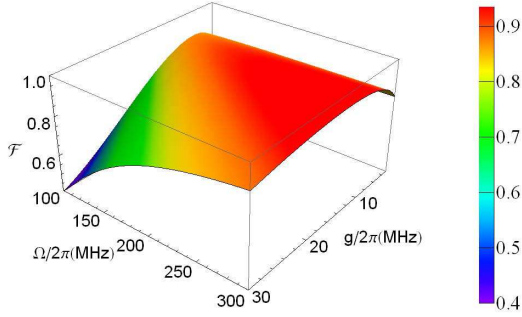} \vspace*{%
-0.08in}
\end{center}
\caption{(color online). Fidelity versus $\Omega/2\protect\pi$ and $g/2%
\protect\pi$. The figure was plotted for $N=6$, $g_{12}/2\protect\pi=0.1g$,
$g_1=0.95g$, and $g_2=g$.}
\label{fig:3}
\end{figure}

For a flux device with the four levels in Fig.~2(b), $g^{\prime
}$ is on the same order of $g$ (or $g_{2}$)$.$ Thus, choose $g^{\prime }=g$
for simplicity. By the numerical test for $N=6$ and $\Omega/(2\pi)=300$ MHz [15],
we find that for $g/\left( 2\pi \right) \leq 15$
MHz [16], when $g_{12}\leq 0.1g$, the effect of inter-cavity crosstalk on
the operation fidelity is negligible. Therefore, set $g_{12}=0.1g$ for Figs.~3 and 4 below. The
condition $g_{12}\leq 0.1g$ can be met as discussed in [11].

Fig.~3 is plotted for $N=6$, showing that $\left\{ {\mathcal{F}%
,\Omega /\left( 2\pi \right) ,g/\left( 2\pi \right) }\right\} $ are: (i)
0.935,\ 300\ MHz,\ 4\ MHz; (ii) 0.916,\ 200\ MHz,\ 10.5\ MHz; and (iii)
0.870,\ 100\ MHz,\ 6.5\ MHz. These results indicate a high fidelity can
be achieved for $N=6$. To further see how the fidelity varies with $N>6,$
Fig.~4 is plotted for $N\leq 10$ and different $\Omega .$ Fig.~4 shows that
for $N=10$, $\left\{ {\mathcal{F},\Omega /\left( 2\pi \right) ,g/\left( 2\pi
\right) }\right\} $ are: (i) 0.864,\ 300\ MHz,\ 9\ MHz; (ii) 0.827,\ 200\
MHz,\ 7\ MHz; and (iii) 0.743,\ 100\ MHz,\ 4\ MHz. The $g$ values here were
obtained by numerically optimizing the coupling constants. These results
indicate a high fidelity can be obtained even for $N=10$.

\begin{figure}[tbp]
\begin{center}
\includegraphics[bb=0 0 896 585, width=7.0 cm, clip]{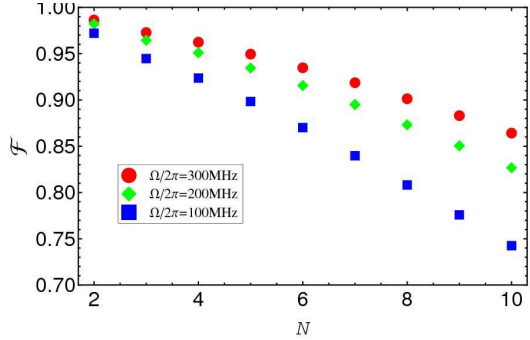} \vspace*{%
-0.08in}
\end{center}
\caption{(color online). Fidelity versus $N$. The figure was plotted for $%
g_{12}/2\protect\pi=0.1g$, $g_1=0.95g$, and $g_2=g$.}
\label{fig:4}
\end{figure}

For cavities $1$ and $2$ with frequencies given above and the $\kappa
_{a_{1}}^{-1}$ and $\kappa _{a_{2}}^{-1}$ used in the numerical calculation,
the required quality factors for the two cavities are $Q_{1}\sim 4.4\times
10^{5}$ and $Q_{2}\sim 6.9\times 10^{5},$ which are readily available in
experiment [17]. Our analysis given here demonstrates that high-fidelity
generation of the NOON states with $N\leq 10$ even for the imperfect device is
possible within the present circuit QED techniques.

This work was supported in part by the National NSFC and BRPC under Grant
Nos.~[11074062, 11374083, 11375003, 11174081, 11034002, 11134003,
2011CB921602], the funds from Hangzhou Normal University under grant
Nos.~[HNUEYT 2011-01-011, HSQK0081, PD13002004], and the funds of Hangzhou
City for the Hangzhou-City Quantum Information and Quantum Optics Innovation
Research Team.

\end{document}